\documentstyle[aps,12pt]{revtex}

\begin{document}
\def\simlt{\stackrel{<}{{}_\sim}}
\def\simgt{\stackrel{>}{{}_\sim}}
\def\cint{\oint}     
\title{Gravitation as a Many Body Problem}
\author{Pawel O. Mazur}
\address{Department of Physics and Astronomy, University of South
Carolina,}
\address{Columbia, SC 29208, USA}
\maketitle
\begin{abstract}
The idea of viewing gravitation as a many body phenomenon is
put forward here. Physical arguments supporting this idea are briefly
reviewed.
The basic mathematical object of the new gravitational mechanics
is a matrix of operators \cite{pom1}.
Striking similarity of the method of R-matrix (QISM)
to the mathematical formulation of the new gravitational mechanics
is pointed out. The s-wave difference Schrodinger equation
describing a process of emission of radiation by a gravitating particle
\cite {pom1} is shown to be analogous to the Baxter equation
of the QISM \cite {skl1}.
\end{abstract}
\vskip 1.2cm
\centerline{\it {Talk given at the Conference
``Beyond the Standard Model V'',}}
\centerline{\it {April 29-May 4 1997, Balholm, Norway.}}

\vspace{1cm}

\section{Introduction}

It is well known that the classical gravitating systems behave in the
way foreign to statistical quantum mechanics. The negative specific heat
of those systems and the phenomenon of gravitational collapse are
different facets of the same reality. The difficulties arise which
necessitate that the complete atomistic description of gravitation be
sought after \cite {pom1,pom2}. An effective spacetime description of
phenomena would be than obtained in the thermodynamical large $N$ limit
\cite {pom2,gorpom}. The well known paradoxes of gravitational physics
which only until recently stood apart are recognized now as intimately
connected \cite {pom1,pom2,gorpom}. They are: {\em i)} an ultimate
gravitational collapse
which also leads to the breakdown of spacetime description;
{\em ii)} a negative specific heat of gravitating systems
and of black holes specifically \cite {bek1,hawk1};
{\em iii)} unitarity nonconservation on black holes \cite
{hawk1,thoof1};
{\em iv)} information loss on black holes, paradox of nonorthodox
causality structure, ``breakdown of physics'' in singularities;
{\em v)} nonrenormalizability of gravitation in the usual scheme of QFT.
The well known fact of nonrenormalizability of gravitation
is not surprising after all because gravitons
are not the right microscopic degrees of freedom.
This is here where the {\em non-collapse postulate}
and the resulting statistical mechanical arguments \cite
{pom2,gorpom} help to
establish that the microscopic degrees of freedom are the Planckian mass
{\em gravitational atoms} rather than gravitons.
Gravitational quanta \cite {pom2} and
gravitons must then appear as collective excitations of a bound state
of many
gravitational atoms. That the microscopic theory of gravitation must
involve new and previously unanticipated degrees of freedom, which have
nothing to do with strings or membranes \cite {suss2,ban}, was first
proposed several years ago \cite {pom1,pom2,lect95}. The idea of
introducing new kinematics, and to be specific, exactly, the
generalization of the Heisenberg matrix mechanics
to the matrix of operators for a gravitating particle was first
described in \cite {pom1} and at the USC Summer 1995 Institute lecture
\cite
{lect95,cn95}. It should be clear that kinematics is fundamental
whereas dynamics depends on a system: there are many Hamiltonians but
only one fundamental relation $[p,q]={h\over {2\pi}i}$. This is why my
proposal for kinematics of the new gravitational mechanics \cite
{pom1} is universal while strings and membranes
are only models \cite{suss2,ban}.

\section{The Fundamental Idea of The New Gravitational Mechanics}

I will briefly present different heuristic arguments which have led
me to the proposal of the new gravitational mechanics \cite {pom1}.
In essence my argument that the fundamental object of gravitational
mechanics is the matrix of operators followed from a careful
examination of the GRT Kepler problem \cite {pom1}. The geodesic motion
is uniformized by a complex$\equiv$symplectic torus. Hence the idea of
a noncommutative torus replacing a symplectic torus has emerged
\cite{lect95,pom1}. Two sets of matrices of operators
${\bf X}^{\mu}$, ${\bf P}_{\mu}$,
which now describe a geodesic motion of a test particle satisfy a matrix
Heisenberg relations \cite {lect95,pom3}.

There exists a subtle
connection between the idea of a matrices of operators describing
a gravitating particle and QISM. The second symplectic structure of the
KdV equation is the same as the Virasoro algebra, the algebra of
classical reparametrization invariance of a geodesic action principle.
Quadratic noncommutative
algebras related to the Yang-Baxter equation are central to the quantum
inverse scattering method (QISM) \cite {skl1}. QISM or R-matrix method
deals directly with a matrix of operators as the fundamental object
\cite {skl1}.

I have noticed long time ago that a classical geodesic motion of a test
particle (massive or massless) in the gravitational field of a
Schwarzschild black hole is related to the single soliton solution of
the KdV equation. This is so because

\begin{equation}
u=r^{-1}={1\over 2KM}({1\over 3}+4{\cal P}(\phi;\omega,\omega')),
\end{equation}

where $K={G\over c^2}$ is the Einsteinian gravitational constant,
and $M$ is the mass of a black hole,
is a solution to the geodesic equation. This motion is described
by the doubly periodic Weierstrass ${\cal P}$-function, with two
periods $2\omega$, $2\omega'$ which determine a lattice
on the complex $\phi$ plane.
The classical geodesic motion in the Schwarzschild black hole
gravitational field is uniformized by a complex torus. I have suggested
that this complex torus be regarded as a symplectic torus. This has
led me to the canonical Sommerfeld phase integral

\begin{equation}
I=\cint p_rdr
\end{equation}

over a homotopically nontrivial contour on an elliptic curve
uniformizing the actual physical motion (double
covering of the complex $r$-plane).
Similar construction can be easily
carried through for the geodesic motion in the Kerr black hole
gravitational field with the effect that

\begin{equation}
I=4{\pi}iG{\delta}{M_C}^2=i{\delta}{A_{bh}\over 4G}=i{\delta}S_{bh} ,
\end{equation}

where $M_C$ is the Christodoulou irreducible mass\footnote{The
Christodoulou irreducible mass is defined by the formula
$M^2={M_C}^2 + {J^2\over 4G^2{M_C}^2}$.} \cite {christo},
$A_{bh}$ is an area of a black hole horizon, and $S_{bh}$ is a black
hole entropy \cite {bek1,hawk1} ($\hbar=c=k=1$ is assumed here).
Semiclassical
quantization conditions of Sommerfeld can be easily extended to the case
of the complexified KAM torus (complexified integrable systems) \cite
{pom1,lect95,pom3}. The two angle variables on a complex uniformizing
torus
lead to the generalized Heisenberg matrix mechanics. It is essential
here that the two periods are distinct in character, as one of them is
complex (purely imaginary for bound orbits). I have followed the basic
idea of transition to matrix mechanics as described by Dirac \cite
{dir,pom1}. According to Dirac \cite {dir} the fundamental idea of
matrix mechanics was to consider {\em two Bohr orbits instead of one}.
Now, two distinct periods in the GRT Kepler problem suggested to me the
idea of considering two sets of two integers related to two periods on
the complex uniformizing torus, and hence a matrix of matrices. This is
the essence of the idea of generalized Heisenberg matrix mechanics
applied to gravitation: a matrix of operators ${\cal O}_{mn}$ as the
fundamental object.
The problem of a particle moving in the gravitational field of two heavy
centers (two extremally charged black holes described classically by the
Majumdar-Papapetrou metric), which is the GRT Pauli problem, shows that
one must now consider an object ${\cal O}_{{\bf mn}}$, where ${\bf
n}$ is a vector multi index; in the Pauli problem this vector
index has only two components. The generalization to other systems is
obvious \cite {pom3}. The action principle for the GRT Kepler and Pauli
problems is the generalization of the geodesic action principle for
the mathematical object ${{\bf X}^{\mu}}_{\bf mn}$:

\begin{equation}
I={1\over {2l^2}}\int Tr({\bf g}_{\mu\nu}({\bf X}){\partial {\bf
X}^{\mu}}{\partial {\bf X}^{\nu}}){d\xi},
\end{equation}

where $l$ is the Planck length, ${\bf g}_{\mu\nu}$
is a metric tensor which is now a matrix of operators,
and ${\partial {\bf X}}={d\over d\xi}{\bf X}$,
with $\xi$ a complex parameter. The trace
operation $Tr({\bf A}{\bf B})$ is defined as a trace over two vector
spaces on which a matrix of matrices (operators) acts.
In this way one recovers in an explicit mathematical form an early
heuristic suggestion of quantization of a black hole area originally due
to Bekenstein \cite {bek2,grg19,thoof2,suss1}. It becomes clear now
that the spacetime picture associated with the surface of a black hole
horizon divided into many Planckian area cells is quite misleading.
Such a spacetime picture, sometimes called {\em the holographic
principle} \cite {bek2,thoof2,suss1}, is equally incorrect as a toy
idea of the noncommutative configuration space would be
at the time of the old quantum theory.
It is the phase space, or symplectic geometry, which
becomes noncommutative upon transition to quantum mechanics.
Physical arguments based on the {\em non-collapse} postulate and on the
s-wave difference Schrodinger equation are compatible. We have found a
spectrum of collective excitations of a system of gravitational atoms
\cite {pom1,pom2,gorpom}.

\section{The Non-Collapse Postulate and Integrable Many Body Systems}

The method of R-matrix or the quantum inverse scattering method (QISM)
has proven quite powerful in the many body problem solving \cite
{skl1}. It is perhaps
not too naive to expect that this method could be also applied to the
phenomenon of the universal gravitation once it is demonstrated that a
gravitational mass behaves as a many body system.

That gravitation is a many body problem follows from the black hole
thermodynamics with the non-collapse hypothesis incorporated explicitly
\cite {pom1,pom2,gorpom}. The Bekenstein entropy \cite {bek1} of a black
hole (and its negative specific heat) does not really imply
that gravitation is actually a many body problem.
Even though invoking the concept of
entropy in the context of black holes implicitly means that some kind
of atomism is assumed, the many body (atomic) nature of a gravitational
mass was not uncovered until only quite recently \cite {pom2,gorpom}.
The concept of gravitational quanta and gravitational atoms has appeared
quite naturally once the non-collapse hypothesis was introduced.
Previously, the information theoretic interpretation of entropy was
adopted in accord with an apparent loss of information inside black
holes \cite {bek1}. Therefore, the concept of gravitational atoms and
gravitational quanta did not appear earlier.
On the other hand elementary planckian
cells of black hole horizon area were almost always mentioned in the
context of black hole thermodynamics \cite
{bek1,bek2,thoof1,thoof2,suss1}. Clearly, it is a long way from
the concept of the black hole horizon area cells to the idea that the
physical nature of a gravitational mass is best described as a many body
problem. The arguments to this effect are of physical character and were
first discovered several years ago \cite {pom1,pom2,grg19}. This is to
say, that our method
is model independent and this conclusion was obtained long before the
so-called D-brane soliton approach to certain unphysical black holes was
presented in the literature \cite {strovaf}.

The central character of
the non-collapse
hypothesis in the demonstration that gravitation is a many body
phenomenon is difficult to overestimate. The {\em gedanken experiment}
presented in the next section was a basis of our hypothesis that the
total specific heat of a quantum black hole is positive
\cite {pom2,gorpom}. This is clearly connected to the problem of
unitarity conservation. With the help of this single postulate we were
able to demonstrate that a gravitational mass-energy behaves as a number
of correlated harmonic oscillators \cite {pom2,gorpom}. This is in a
complete agreement with the s-wave difference Schrodinger equation
presented in \cite {pom1}. I have called these correlated harmonic
oscillators {\em gravitational quanta} because they are related to the
gravitational mass.

Now, it became clear to me that the simplest properties of gravitational
quanta (and gravitational masses) suggest that they are weakly coupled
for a large total mass-energy $E$ and that the total mass-energy of this
system is a sum of energies of collective excitations

\begin{equation}
E=\sum_{n} {\epsilon}_{n}.
\end{equation}

In particular, for a large natural number $N$ a typical
energy scale for collective excitations is

\begin{equation}
{\epsilon}_{n}\sim {{\mu}\over \sqrt{N}},
\end{equation}

where $\mu$ is a Planck mass. The total
mass-energy scales with $N$ as follows:

\begin{equation}
E\sim {\mu}\sqrt{N}.
\end{equation}

This is a general property of integrable quantum many body systems
that a total energy is a sum over energies of collective excitations
\cite {skl1}.
In my previous work on the collective excitations of a black hole I have
proposed an s-wave difference Schrodinger wave equation which describes
energy levels of a black hole \cite {pom1}. I have taken a null shell
model for
the s-wave scalar particle mode in the geometric optics approximation,
which
is only adequate near a black hole horizon, and I have quantized the
collective degree of freedom of such a shell. The energy spectrum came
out a harmonic oscillator spectrum. Such a procedure is not quite
unique, as the more detailed analysis has shown. On the other hand
the validity of such
a model is limited only to the region near a black hole horizon, and as
such this model should be considered an approximation. One can
conclude from
this that membrane models are not to be considered valid unless
qualitative agreement is reached with other independent
arguments of physical nature as the one presented above.
We should keep in mind that we move on a new
ground here and we should seek the consistency of arguments.
The formal analogy which comes to mind is based on two observations:

{\em{ i)} Gravitation is a many body phenomenon;
the total mass-energy of a gravitating mass is a sum of energies of
collective excitations.}

{\em{ii)} Quantum integrable models of many body systems are leading
to the same property of the total energy, and in addition to that the
method of separation of variables (SOV) \cite {skl1} applied to
such systems leads to the Schrodinger wave equation,
called in this case the
Baxter equation, which, in general, is a difference equation.}

Our basic observation is that the s-wave Schrodinger equation
describing collective excitations of a black hole \cite {pom1}:

\begin{equation}
x(\psi(x+iL)+\psi(x-iL))=(x+2{\epsilon})\psi(x) ,
\end{equation}

where $\epsilon$ is an energy of a collective excitation
and $L\sim {1\over M}$ in Planck units,
is of the same type as Baxter equations for collective excitations of
integrable quantum many body systems \cite {skl1}. The question which
occurs naturally now is the following:
What kind of mathematical structure describing a gravitating particle
leads to the difference Schrodinger equation presented in \cite {pom1} ?
If we look closer into the derivation of the Baxter equations in the
QISM approach of Sklyanin \cite {skl1}, we also notice that the
basic mathematical structure is the matrix of operators constrained by
some
quadratic relations. Associativity of such quadratic algebras leads to
the Yang-Baxter equations (YBE). Now, my idea of the second Heisenberg
algebra related to the presence of the second period of Nature
(the second angle variable on a complex torus) in the
GRT Kepler problem \cite {pom1} was to consider a matrix of operators as
an observable in the new gravitational mechanics \cite {pom1,pom3}.
This observation has a profound meaning, and it cannot
be just an accident, as it is based on physical arguments.
One must follow the lead which this connection has opened up \cite
{pom3}.

\section{The Gedanken Experiment and the Non-Collapse Hypothesis}

Consider the following situation. A large number of identical
gravitating particles (mini black holes), which for simplicity will be
assumed to be spinless, are contained in a large box. This large box can
be considered to be filled up with a black body radiation
at some temperature $T$. This cavity black body radiation
is considered to be in thermodynamical equilibrium with a cavity walls
which
are also kept at a constant temperature $T$. The next step is to assume
that a black body radiation is in statistical equilibrium with mini
black holes. Unless small identical black holes
are in statistical equilibrium among themselves due to collision
processes the second law of thermodynamics is violated.
The {\em non-collapse hypothesis} can be concisely formulated as the
statement
to the effect that in the processes of collisions between identical mini
black holes inelastic processes do not happen (hence non-collapse), or
that collisions are elastic with a high probability. This is quite
similar to the situation considered once by Einstein \cite {ein1}.
Our gedanken experiment is concerned with identical
gravitating particles in thermodynamical equilibrium, which is, on the
other hand, impossible according to general relativity (GRT). In this
situation GRT predicts a merger of two or more mini black holes in
the process of a head-on collision. The phenomenon of gravitational
collapse precludes the possibility of statistical equilibrium to be
achieved in processes of mini black hole collisions. Hence, the
hypothesis of statistical equilibrium in our gedanken experiment
is a qualitatively new element introduced into the physical theory.
Gravitational collapse
in the head-on collisions of mini black holes does not happen. If it
does happen, as GRT predicts, then the second law of thermodynamics
is not valid in the situation described by our gedanken experiment.

The classic argument of Einstein \cite {ein1} applied by him to a
classical
gas of atoms in statistical equilibrium with a black body radiation
and among themselves leads to the Planck distribution
formula for radiation once the Bohr formula $E_{m}-E_{n}=h{\nu}_{mn}$
(and the postulate of spontaneous and stimulated emission of radiation)
was assumed for the processes of absorption and emission of radiation
by atoms. Reversing this classic argument, when applied to our
gedanken experiment, must necessarily lead to the quantized mass-energy
levels of mini black holes \cite{bek2,grg19,pom1,pom2,gorpom}.
The new postulate was made in derivation of this conclusion. This is the
non-collapse postulate which logically plays the same role in our
gedanken experiment as the postulate of spontaneous emission of
radiation has played in the derivation of Planck formula \cite {ein1}.
In our case we postulate the Planck formula but we want to argue for the
Bohr formula. We have thus reached the conclusion that a gravitating
mass behaves statistically as if it consisted of a huge number of
gravitational quanta. These quanta could be identified with collective
excitation of a large number of Planckian mass scale gravitational
atoms \cite{pom2,gorpom}. Statistical considerations are qualitative by
nature and they do not lead to the definitive mathematical models for
the dynamics of those gravitational atoms. The only important statement
which follows from our work \cite {pom2,gorpom} is the definitive
argument about the physical nature of gravitating particles. The nature
of a gravitational mass is now understood qualitatively. Einsteinian
gravitation is by its nature the many body phenomenon.

\vspace{1mm}

This work was supported by the NSF grant to the University of South
Carolina.
\vspace{1mm}

{\em Acknowledgements.}
I wish to thank organizers of BSMV for their kind hospitality during
the conference.
\vspace{-0.5cm}


\begin{references}
\vspace{-1.4cm}
\bibitem{pom1}    P. O. Mazur, Acta Phys. Polon. {\bf 26}, 1685 (1995).
\bibitem{skl1}    E. K. Sklyanin, in {\em Quantum Group and Quantum
Integrable Systems}, pp. 63-97, ed. Mo-Lin Ge, World Scientific,
Singapore,
1992.
\bibitem{dir}     P. A. M. Dirac, in {\em Directions in Physics,
Development of Quantum Mechanics}, pp. 2-20, especially pp. 4-5, John
Wiley and Sons, New York 1978.
\bibitem{pom2}    P. O. Mazur, Acta Phys. Polon. {\bf 27}, 1849 (1996).
\bibitem{gorpom}  A. Z. Gorski and P. O. Mazur, hep-th/{\bf 9704179},
submitted to Phys. Rev. Lett. (1997).
\bibitem{bek1}    J. D. Bekenstein, Phys. Rev. {\bf D7}, 2333 (1973).
\bibitem{hawk1}   S. W. Hawking, Commun. Math. Phys. {\bf 43}, 199 (1975).
\bibitem{thoof1}  G. `t Hooft, Nucl. Phys. {\bf B335}, 138 (1990).
\bibitem{lect95}  P. O. Mazur, lecture at the USC Summer Institute,
July 1995.
\bibitem{pom3}    P. O. Mazur, {\em On the Fundamental Idea of the New
Gravitational Mechanics}, unpublished (1996).
\bibitem{christo} D. Christodoulou, Phys. Rev. Lett. {\bf 25},
1596 (1970).
\bibitem{bek2}    J. D. Bekenstein, Lett. Nuovo Cimento {\bf 11},
467 (1974).
\bibitem{grg19}   P. O. Mazur, GRG {\bf 19}, 1173 (1988).
\bibitem{thoof2}  G. `t Hooft, in {\em Salamfest},
gr-qc/{\bf 9310026}, (1993).
\bibitem{suss1}   L. Susskind, J. Math. Phys. {\bf 36}, 6377 (1995).
\bibitem{strovaf} A. Strominger and C. Vafa, Phys. Lett. {\bf 379},
99 (1996).
\bibitem{ein1}    A. Einstein, Phys. Z. {\bf 18}, 121 (1917).
\bibitem{cn95}    A. Casher and S. Nussinov, hep-ph/{\bf 9510364},
(1995).
\bibitem{suss2}   T. Banks, W. Fischler, S. H. Shenker and L. Susskind,
Phys. Rev. {\bf D55}, 5112 (1997).
\bibitem{ban}     T. Banks, hep-th/{\bf 9706168}, (1997).

\end{references}
\end{document}